\documentstyle[psfig,prb,aps]{revtex}
\begin{document}
\draft

\twocolumn[\hsize\textwidth\columnwidth\hsize\csname
@twocolumnfalse\endcsname

\title{
Melting in large sodium clusters: An orbital-free molecular dynamics
study. 
}
\author{Andr\'es Aguado, Jos\'e M. L\'opez, and Julio A. Alonso}
\address{Departamento de F\'\i sica Te\'orica,
Universidad de Valladolid, Valladolid 47011, Spain}
\author{M. J. Stott}
\address{Department of Physics, Queen's University, Kingston,
Ontario K7L 3N6, Canada}
\maketitle
\begin{abstract}
The melting-like transition in sodium clusters
Na$_N$, with N=55, 92, and 142 is studied by using
constant-energy molecular dynamics simulations. An orbital-free version
of the Car-Parrinello technique is used which scales 
linearly with system size allowing investigation of the 
thermal behaviour of large clusters. The ground state isomer of Na$_{142}$
(an uncomplete three-shell icosahedron) melts in two steps: the first one
(at $\approx$ 240 K) is characterized by the high mobility of the atoms
located on the cluster surface;  the second, homogeneous melting (at
$\approx$ 270 K), involves diffusive motion of all the atoms across the cluster.
For the case of Na$_{92}$, the icosahedral structure has a larger number of
surface vacancies, and melts in two well separated steps, 
surface melting at $\approx$ 130 K and homogeneous melting at $\approx$ 240 K.
Na$_{55}$, a complete two-shell icosahedron,
melts in a single stage at $\approx$ 190 K.
Our results on homogeneous melting for Na$_{142}$ and Na$_{92}$
are in excellent agreement with recent experimental determinations of
melting temperatures and latent heats. However, the experimentally observed
enhancement of the melting temperature around N=55 is not reproduced by the
calculations.
\end{abstract}

\vskip2pc]

\section{Introduction}

The melting-like transition in finite clusters consisting of a small number of
atoms
is of fundamental interest as clusters are 
often produced in a disordered "liquid" state,\cite{Ell95} and it is also
relevant to applications of clusters. For example, the catalytic activity
of small platinum clusters depends critically on their melting temperatures.
\cite{Wan98} Recent experimental advances reveal some
details of the melting-like transition but, at the same time, show new and
interesting features. Martin \cite{Mar96} determined the
cluster size dependence of the melting temperature T$_m$ of large
sodium clusters, composed of thousands of atoms, by observing the vanishing of
the atomic shell structure in the mass spectra upon heating. It was 
concluded that T$_m$ grows with cluster size, but the results did not
extrapolate yet to the T$_m$ of the bulk. Peters {\em et al.} \cite{Pet98}
performed X-ray diffraction experiments on large (50 nm)
Pb clusters and observed the
occurrence of surface melting before homogeneous melting.  Electron diffraction
\cite{Mai99} may also help in detecting a surface melting stage.
\cite{Wan98} Haberland and coworkers \cite{Sch97} have studied the
variation with temperature of the photofragmentation spectra of Na$_N$
(N=50--200) clusters, and have deduced the melting temperatures of
the clusters. They find that for some cluster sizes the 
melting temperature is a local maximum not in exact correspondence with 
either electronic or atomic shell closing numbers, but
bracketed by the two, suggesting that both effects are relevant to 
the melting process.

A number of computer simulations of melting in small metallic and nonmetallic
clusters has been reported, the majority of which employed phenomenological 
interatomic potentials.\cite{Jel86,Cle98,Cal99} The use of such parameterized
potentials allows the consideration of long dynamical trajectories for large
clusters.\cite{Cle98,Cal99} {\em Ab initio} methods,\cite{Car85}
which have also been 
used, accurately treat the electronic structure of the cluster,
but are much more expensive computationally and are usually 
restricted to the study of small clusters for short dynamical trajectories.
\cite{Rot91} Recently, Rytk\"onen {\em et al.} \cite{Ryt98} have performed
{\em ab initio} molecular dynamics (aiMD) simulations of the melting of a
sodium cluster with 40 atoms, but such a large cluster required the 
use of a fast heating rate. These aiMD treatments use the Kohn-Sham 
(KS) form\cite{Koh65} of density functional theory (DFT), and 
orthogonalization of the one-electron KS orbitals is the limiting step in 
their performance.  However, DFT shows that the total energy of
the electronic system can be expressed in terms of just the
electronic density,\cite{Hoh64} and orbital-free (OF) versions of the aiMD
technique based
on the electron density have been developed and employed, both in solid 
state \cite{Pea93,Gov99} and cluster \cite{Sha94,Gov95,Bla97,Agu99} 
applications. These OF methods scale linearly with the system size allowing 
the study of larger clusters for longer simulation times 
than typical aiMD simulations. However, quantum shell effects
are neglected, so that features associated with electronic shell
closings are not reproduced.

Previously,\cite{Agu99} we have used the orbital-free molecular 
dynamics (OFMD) method to study the melting process in small sodium clusters,
Na$_8$ and Na$_{20}$, clusters outside the range covered by Haberland's 
photofragmentation
experiments.\cite{Sch97} Here, we report constant energy OFMD
simulations in a study of the melting-like transition in larger clusters,
Na$_{55}$, Na$_{92}$ and Na$_{142}$, which are within the size range 
covered in those experiments, and for which a full {\em ab initio} treatment
of their thermal properties would be impractical. Even for the OFMD method
those large clusters represent a very substantial computational effort.
The aim of our work is 
to study the mechanisms by which the melting-like transition proceeds in 
these large clusters. 
In the next section we briefly present some technical details of the method. 
The results are presented and discussed in section III and, finally, 
section IV summarizes our main conclusions.

\section{Theory}

The orbital-free molecular dynamics method is a Car-Parrinello total
energy scheme\cite{Car85}
which uses an explicit kinetic-energy functional of the 
electron density, and has the electron
density as the dynamical variable, as opposed to the KS single particle
wavefunctions. In contrast to simulations which use empirical interatomic
potentials, the detailed electronic structure and the electronic contribution
to the energy and the forces on the ions are recalculated efficiently every
atomic time-step. The main features of the energy functional and
the calculational scheme have been described at length in previous work,
\cite{Pea93,Sha94,Bla97,Agu99} and details of our method are as
described by Aguado et al.\cite{Agu99} In brief, the electronic kinetic 
energy functional of the electron density, $n(\vec r)$, corresponds to 
the gradient expansion around the homogeneous limit through second order 
\cite{Hoh64,Mar83,Yan86,Per92}
\begin{equation}
T_s[n] =
T^{TF}[n] + {\frac{1}{9}} T^W[n],
\end{equation}
where the first term is the Thomas-Fermi functional 
\begin{equation}
T^{TF}[n] = \frac{3}{10}(3\pi^2)^{2/3}\int n(\vec r)^{5/3}d\vec r,
\end{equation}
and the second is the lowest order gradient correction, where T$^W$,
the von Weizs\"acker term, is given by 
\begin{equation}
T^{W}[n] = \frac{1}{8}\int \frac{\mid \nabla n(\vec r) \mid^2}{n(\vec r)}d\vec r.
\end{equation}
The local density approximation is used for exchange and correlation.\cite
{Per81,Cep80} In the external field acting on the electrons,
$V_{ext}(\vec r) = \sum_n v(\vec r -\vec R_n)$, we take $v$ to be
the local pseudopotential of Fiolhais {\em et al.}, \cite{Fio95} which
reproduces well the properties of bulk sodium and has been shown to have good
transferability to sodium clusters.\cite{Nog96}

The cluster is placed in a unit cell of a cubic superlattice,
and the set of plane waves periodic in the superlattice
is used as a basis set to expand the valence density.
Following Car and Parrinello,\cite{Car85} the coefficients
of that expansion are regarded as generalized coordinates of a set of 
fictitious classical particles, and the corresponding Lagrange equations of 
motion for the ions and the electron density distribution are solved as 
described in Ref. \onlinecite{Agu99}.

The calculations for Na$_{92}$ and Na$_{142}$ used a supercell of edge 
71 a.u. and the energy cut-off in the plane wave expansion of the 
density was 8 Ryd. For Na$_{55}$, the cell edge was 64 a.u. and the 
energy cut-off 10 Ryd. In all cases, a 64$\times$64$\times$64 grid was used. 
Previous tests \cite{Agu99} indicate that the cut-offs used give good
convergence of bond lengths and binding energies. The fictitious
mass associated with the electron density coefficients
ranged between 1.0$\times$10$^8$ and 3.3$\times$10$^8$ a.u.,
and the equations of motion were integrated using the Verlet 
algorithm \cite{Ver65} for both electrons and ions with a time step 
ranging from $\Delta$t = 0.73 $\times$ 10$^{-15}$ sec. for the simulations 
performed at the lowest temperatures, to $\Delta$t = 0.34 $\times$ 
10$^{-15}$ sec. for those at the highest ones. These choices resulted in 
a conservation of the total energy better than 0.1 \%. 

The first step of the simulations
was the determination of low temperature isomers for each of the three cluster
sizes. For such large clusters it is very difficult to find
the global minimum because the number of different local
minima increases exponentially with the number of atoms in the
cluster. Instead, one has to adopt structures that are likely to have the
main characteristics of the ground state. 
We have assumed icosahedral growth. Thus, for Na$_{142}$ we removed 
five atoms from the surface of a 147 atom three-shell perfect icosahedron.
For Na$_{92}$, we constructed an icosahedral isomer by following the 
growing sequence described by Montejano-Carrizales {\em et al,} \cite{Mon96}
and for Na$_{55}$ we took a perfect two-shell icosahedron. 
We have also used dynamical simulated annealing,\cite{Car85} to generate
low temperature isomers, but this procedure always led to amorphous structures 
for Na$_{92}$ and Na$_{142}$, and to a nearly icosahedral structure for 
Na$_{55}$. 

Several molecular dynamics simulation runs at different constant energies 
were performed in order to obtain the caloric curve for each icosahedral
isomer.
The initial positions of the atoms for the first run were taken by slightly 
deforming the equilibrium low temperature geometry of the isomer.
The final configuration of each run served as the starting geometry for the
next run at a different energy. The initial velocities for every new run were
obtained by scaling the final velocities of the preceding run. The total 
simulation times varied between 8 and 18 ps for each run at constant energy.

A number of indicators to locate the melting-like transition were employed.
Namely, the specific heat defined by \cite{Sug91}
\begin{equation}
C_v = [N - N(1-\frac{2}{3N-6})<E_{kin}>_t ~ <E_{kin}^{-1}>_t]^{-1},
\end{equation} 
where N is the number of atoms and $<>_t$ indicates the average along a
trajectory; the diffusion coefficient, 
\begin{equation}
D = \frac{1}{6}\frac{d}{dt}<r^2(t)>,
\end{equation} 
which is obtained from the long time
behaviour of the mean square displacement 
$<r^2(t)> = \frac{1}{Nn_t}\sum_{j=1}^{n_t}\sum_{i=1}^N [\vec R_i(t_{0_j} + t) -
\vec R_i(t_{0_j})]^2$, where $n_t$ is the number of time origins, $t_{0_j}$,
considered along a trajectory;
the time evolution of the distance between each atom
and the center of mass of the cluster 
\begin{equation}
r_i(t) = \mid \vec R_i(t) - \vec R_{cm}(t)\mid,
\end{equation} 
and finally, the radial atomic density,
averaged over a whole dynamical trajectory,
\begin{equation}
\rho (r) = \frac{dN_{at}(r)/dr}{4\pi r^2}, 
\end{equation} 
where $dN_{at}(r)$ is the number of
atoms at distances from the center of mass between r and r + dr.

\section{Results}

The lowest energy structure of sodium clusters of medium size is not known.
DFT calculations for Na$_{55}$ performed by K\"ummel {\em et al.}\cite{Kum99}
using an approximate structural model (CAPS model, where the total
pseudopotential of the ionic skeleton is cylindrically averaged)\cite{Mon95}
give a structure close to icosahedral. Near-threshold photoionization mass
spectrometry experiments suggest icosahedral structures for large sodium
clusters with more than 1400 atoms,\cite{Mar91} so incomplete icosahedral
structures are plausible candidates for Na$_{92}$ and Na$_{142}$. For
this reason we have adopted for
Na$_{142}$ an isomer obtained by removing five atoms from a 
perfect three-shell icosahedron. The icosahedral growing sequence in nickel 
clusters has been studied by Montejano-Carrizales {\em et al,}\cite{Mon96} 
who have shown that the 12 vertices of the outermost shell are the 
last sites to be occupied. Assuming the same growth sequence for sodium
clusters, we have removed five atoms from the vertex positions of Na$_{147}$,
testing all possibilities, and have then relaxed the resulting 
structures. In the most stable structure thus formed, the five vacancies 
form a pentagon. For Na$_{92}$, we have adopted the umbrella growing model 
of Montejano-Carrizales {\em et al.} \cite{Mon96} The resulting structure
corresponds to three complete umbrellas capping a
Na$_{55}$ icosahedron. Low-temperature dynamical 
trajectories verify that these structures are indeed stable isomers of 
Na$_{92}$ and Na$_{142}$. The icosahedral isomers are more stable than the 
lowest energy amorphous isomers which were found by simulated annealing
(0.017 eV/atom and 0.020 eV/atom for Na$_{92}$ and Na$_{142}$ respectively).
Calvo and Spiegelmann have studied sodium
clusters in the same size range, using pair potential and tight-binding (TB)
calculations, \cite{Cal99} and have also predicted 
icosahedral structures for Na$_{55}$,
Na$_{93}$, Na$_{139}$ and Na$_{147}$.

For each icosahedral cluster we have calculated the so-called caloric curve
which is the internal temperature as a function of the total energy, where
the internal temperature is defined as the average of the ionic kinetic 
energy \cite{Sug91}.
A thermal phase transition 
is indicated in the caloric curve by a change of slope, the slope being the 
specific heat; the height of the step gives an estimate of the latent heat 
of fusion.
However, melting processes are more easily recognised as peaks in the specific 
heat as a function of temperature. These have been 
calculated directly from eq. (4) 
and are shown together with the caloric curves
in figures 1--3. The specific heat peaks occur at the same
temperatures as the slope changes of the caloric curve giving
us confidence in the convergence of our results as the two quantities have
been obtained in different ways.

The specific heat curve for Na$_{142}$ (fig. 1) displays two main
peaks at T$_s \approx$ 240 K and T$_m \approx$ 270 K, suggesting a two 
step melting process, but these are so close
that only one slope change in the caloric curve can be distinguished.
Our analysis below shows that homogeneous melting occurs at T$_m 
\approx$ 270 K
in excellent agreement with the experiments of 
Haberland and coworkers,\cite{Sch97} who give an approximate value of 280 K. 
The latent heat of fusion estimated from the step at T$_m$ in the caloric curve
is q$_m \approx$ 15 meV/atom, again in good agreement
with the experimental value of $\sim$ 14 meV/atom. However, the 
premelting stage at T=T$_s$ is not detected in the experiments, but our results 
are not inconsistent with this because the two calculated
peaks in the specific heat are close together and the height of the first 
peak is much smaller than that of the second; consequently they could be difficult to
distinguish experimentally. Calvo and Spiegelmann \cite{Cal99}
have performed Monte Carlo (MC) simulations using a semiempirical
many-atom potential, and the lowest-energy isomer they found for Na$_{139}$ was 
also an incomplete three-shell icosahedron, in this case with 8 surface vacancies. 
They also report two close peaks in the specific heat curve indicating a 
two-step melting process, with T$_s \approx$ 210 K and T$_m \approx$ 230 K.
They concluded that these two temperatures become closer as the cluster size 
increases, so that for clusters with more than about 100 atoms
there is effectively just one peak in the specific heat and a single-step melting.
Tight binding (TB) molecular dynamics calculations
were performed by the same authors.\cite{Cal99} Although the
melting temperatures were found to be different from
those obtained with the semiempirical potentials (TB tends to overestimate the
experimental values, while empirical potentials tend to underestimate them), the
qualitative picture of melting in two close steps was the same.

The results for Na$_{92}$ are shown in fig. 2.
Two-step melting is again observed, with a small
prepeak in the specific heat at 
T$_s \approx$ 130 K and a large peak, corresponding to
homogeneous melting, at T$_m \approx$ 240 K. In this
case T$_s$ and T$_m$ are well separated, but the first peak is much smaller
than the second, which could again account for the absence of the prepeak in
the experiments. The calculated temperature and latent heat for the homogeneous 
melting stage, T$_m \approx$ 240 K and q$_m \approx$ 8 meV/atom, are again
in excellent agreement with the experimental values,\cite{Sch97} 250 K and 
7 meV/atom respectively. Calvo and Spiegelmann\cite{Cal99} arrive at similar
conclusions based on MC
simulations for Na$_{93}$ using phenomenological potentials:
a small bump near
100 K and a main peak near 180 K. Their TB simulations give values for those two
temperatures roughly 100 K higher.

The experiments \cite{Sch97} indicate a substantial enhancement of the 
melting temperature in the size region around N=55 atoms. The reported 
melting temperature of Na$^+_{55}$ is 325 K, surprisingly higher than that 
of Na$_{142}^+$, which is a local maximum in the size region of the 
third icosahedral shell closing. Our simulations do not reproduce this
enhancement of T$_m$ for Na$_{55}$
and predict that this cluster melts in a single stage at 
T$_m \approx$ 190 K (fig. 3), a result found also by Calvo and Spiegelmann.
\cite{Cal99} The OFMD method does not account for electronic quantum-shell 
effects, and full KS calculations may be needed to clarify this 
discrepancy, although it is not clear {\em a priori} how 
electronic shell effects could 
shift the value of T$_m$ by such a large amount. Of course, another possibility
is that the icosahedron is not the ground state isomer. 
However, K\"ummel {\em et
al}\cite{Kum99} have recently found that the experimental photoabsorption
spectrum of Na$_{55}^+$ is best reproduced with a slightly oblate isomer which
is close to icosahedral. We have also investigated a bcc-like growing 
sequence finding that bcc structures are less stable than icosahedral ones 
for all cluster sizes studied. Also, we studied the melting behavior 
of a Na$_{55}$ isomer with bcc structure and did not find an enhanced melting 
temperature for it either. In summary, the discrepancy between experiment and
theory for Na$_{55}$ deserves further attention.

Various quantities have been analyzed in order to investigate the nature
of the transitions at T$_s$ and T$_m$. 
The short-time averages (sta) of the distances between each atom and the 
center of mass of the cluster, $<r_i(t)>_{sta}$, have been 
calculated, and the cluster evolution during the trajectories has been 
followed visually using computer graphics. The $<r_i(t)>_{sta}$ curves 
for Na$_{142}$ are 
presented in Figs. 4-6  for three representative temperatures. At low 
temperatures (Fig. 4) the values of $<r_i(t)>_{sta}$ are almost independent 
of time.
The movies show that the clusters are solid, the atoms 
merely vibrating around their equilibrium positions. Curve crossings are due to
oscillatory motion and slight structural relaxations rather than 
diffusive motion. At this low temperature quasidegenerate groups which are 
characteristic of the symmetry can be distinguished: one line near the centre of mass 
of the cluster identifies the central atom (its position does not exactly 
coincide with the center of mass because of the location of the five surface 
vacancies); 12 lines correspond to the first icosahedral shell; another 42
complete the second shell, within which we can distinguish the 12 vertex atoms 
from the rest because their distances to the centre of mass are slightly 
larger; finally, 82 lines describe the external shell, where again we 
can distinguish the 7 vertex atoms from the rest.
 
The radial atomic density distributions with
respect to the cluster center, $\rho (r)$, are shown for Na$_{142}$
in Fig. 7.
At the lowest temperature, T=30 K, the atoms in the icosohedral isomer
are distributed in three well separated shells, a surface shell and two inner
shells; as discussed above, subshells form in the second and third shells.
The shell structure is still present at T=130 K.

Figures 5 and 7 show that at T=160K the
atomic shells of the Na$_{142}$ cluster are still well defined, but the
movies reveal isomerization transitions, similar to those found 
at the begining of the melting-like transition of Na$_8$ and Na$_{20}$,
\cite{Agu99} with no true diffusion.
These isomerizations involve the motion of vacancies 
in the outer shell, in such a way that different 
isomers are visited which preserve the icosahedral structure. 
The onset of this motion is gradual and does not lead to features in the 
specific heat, although it is detected in the temperature evolution of the
diffusion coefficient (see Fig. 8 and discussion below). 
The true surface melting stage does not develop in the icosahedral
isomer until a temperature of T$_s\approx$ 240 K is reached. 

Fig. 6 shows the time evolution of $<r_i>_{sta}$ for Na$_{142}$ at a temperature 
T= 361 K at which the cluster is liquid with all the atoms diffusing throughout 
the cluster. Some specific cases of atoms that 
at the begining of the simulation are near (far from) the center of mass of 
the cluster and end in a position far from (near) the center of mass of the
cluster are shown in boldface. The atomic
density distribution at 280 K,
a temperature just above the melting point, 
is nearly uniform across the cluster, a radial expansion
of the cluster by about 5 bohr units 
is evident,
and the surface is more diffuse.

The $<r_i(t)>_{sta}$ curves for Na$_{92}$ at low temperature are
qualitatively similar to those of Na$_{142}$.
Na$_{92}$ shows surface melting at T$_s\approx$ 130 K.
This temperature is in the range where the isomerization processes in Na$_{142}$
set in, but the larger number of vacancies in the surface shell of Na$_{92}$
allows for more rapid surface diffusion and these processes give rise to a distinct
peak in the specific heat.

Na$_{55}$ is a perfect two-shell icosahedron, so 
surface atoms have no empty sites available to move to, and diffusion within 
an atomic shell is almost
as difficult as diffusion across different shells. When the
surface atoms have enough energy to exchange positions with one another 
they can as easily migrate throughout the whole cluster, and melting 
proceeds in a single stage at 190 K. Calvo and Spiegelmann\cite{Cal99} have
suggested that this one-step melting is
associated with 
a large energy gap between the ground state icosahedral structure
and the closest low-lying isomers, but this cannot be a general result
for perfect icosahedral metallic clusters, as the details of melting
are material dependent. For example, a surface melting stage has been observed 
in simulations of icosahedral Ni$_{55}$.\cite{Guv93}

The variation of the diffusion coefficient with
temperature is shown in Fig. 8 for Na$_{142}$. 
At temperatures less than about 140K, D is close to zero,
indicating only an oscillatory motion of the atoms.
For temperatures between 140K and T$_s$,
the diffusion coefficient increases indicating that
the atoms in the cluster are not undergoing simple vibrational motion;
the atomic motions are, nevertheless, of the special kind discussed above
that preserve the icosahedral structure. The slope of D(T) increases
appreciably at T$_s$ when surface melting occurs, but there is no noticeable
feature when the cluster finally melts at T$_m$.
The features of D(T) for Na$_{92}$ which are not shown here, are
very similar: D(T) is very sensitive to the surface melting
stage, where appreciable diffusive motion begins, and the homogeneous melting
transition is masked by that effect. We conclude that
the D(T) curve is a good indicator of homogeneous melting only in those cases
where the surface melting stage is absent, as for example in Na$_{55}$.

Our results suggest that the melting transition in 
large icosahedral sodium clusters occurs in a smaller temperature range 
than for small clusters 
such as Na$_8$ or Na$_{20}$,\cite{Agu99} at least near an icosahedral shell
closing. Furthermore, the size of any prepeak 
diminishes with respect to the main homogenous melting peak as the cluster size
increases, that is as the fraction of atoms that
can take part in premelting decreases. Consequently, a homogeneous melting
temperature can be defined with less ambiguity for large clusters.
These comments apply to the caloric and the specific heat curves, which are the
quantities amenable to experimental measurement. In contrast, microscopic 
quantities such as the diffusion coefficient D or the $<r_i(t)>_{sta}$ curves 
are very sensitive to any small reorganization in the atomic arrangement, and
it is difficult to determine the melting transition from the variation of 
these quantities with temperature. 
A helpful structural, as opposed to thermal, indicator of
the melting transition in medium-sized or large clusters is the shape of the
radial atomic density distribution. 
The atomic density displays pronounced shell structure 
at low temperatures which is smoothed out at intermediate temperatures
where the vacancy diffusion and/or surface melting mechanisms are present.
Above T$_m$ the density is flat.

In figure 9 we compare the calculated values of the melting temperature
with the experimental values. Our earlier results for 
Na$_8$ and Na$_{20}$ \cite{Agu99} are also included, although for such small 
sizes there is some ambiguity in defining a melting temperature. 
There is excellent agreement with the experimental
results for Na$_{92}$ and Na$_{142}$.\cite{Sch97} Measurements of
the temperature dependence of the photoabsorption cross sections for Na$_n^+$ 
(n=4--16) have recently been reported.\cite{Sch99,Hab99}
Although the spectra do not show evidence of a sharp melting
transition, some encouraging comparison
between theory and experiment can be made. The 
experimental spectra do not change appreciably
upon increasing the cluster temperature, until at T=105 K (the value given as
the experimental melting temperature of Na$_8$ in Fig. 9) the spectra
begin to evolve in a continuous way. In our study of the melting behaviour of 
Na$_8$ \cite{Agu99} we found a broad transition starting at T=110 K and 
continuing until T=220 K, 
at which point the ``liquid'' state was fully developed. 
This may explain the absence of abrupt changes in the experimental
photoabsorption spectrum
with temperature. In any case, we feel that the good agreement between theory 
and experiment may extend to the small sizes. 
However, our method is not expected
to give accurate results if oscillations in the melting temperature with 
cluster size arise as a consequence of electronic shell effects, which is not
yet known. The discrepancy
for Na$_{55}$ remains intriguing. In this regard it is noteworthy 
that our calculated melting temperatures for the three large clusters fit 
precisely the expected large N behaviour,
T$_m$(Na$_N$)=T$_m$(bulk) + C/N$^{\frac{2} {3}}$, where $C$ is a constant,
and yield as a bulk melting
temperature T$_m$(bulk)=350 K, which is close to the observed value of 371 K. 
A similar extrapolation to the bulk melting temperature 
is not evident in the experimental data.

As it is well known that the specific isomer used to start the heating dynamics
can affect the details of the melting transition, \cite{Bon97} we have also 
studied the melting of amorphous Na$_{92}$ and Na$_{142}$ clusters that we
obtained by a simulated annealing.  For the amorphous Na$_{142}$ cluster 
bulk melting occurred at the same temperature, T=270 K, as for the 
icosahedral cluster, while surface melting took place at a much lower
temperature, T=130 K. However, melting of the amorphous Na$_{92}$ cluster 
took place over a wide temperature, and no sharp transitions were detected.
We attach little significance to these results as the initial structures and
the melting behaviour  must depend on the details of the annealing,
and the subsequent heating.

\section{Discussion and conclusions}

A few comments regarding the quality of the simulations 
are perhaps in order here.
The orbital-free representation of the atomic interactions, although much more
efficient than the more accurate KS treatments, is still
substantially more expensive
computationally than a simulation using
phenomenological many-body potentials. Such potentials contain
a number of parameters that are usually chosen by fitting some bulk and/or
molecular properties. In contrast
our model is free of external parameters, although there are
approximations in the kinetic and
exchange-correlation functionals. 
The orbital-free scheme accounts, albeit approximately, for the effects of the
detailed electronic distribution on the total energy and the forces on the ions.
We feel that this is particularly important in metallic clusters for which a
large proportion of atoms are on the surface and experience a very different
electronic environment than an atom in the interior. Furthermore, the
adjustment of the electronic structure and consequently the energy and forces
to rearrangements of the ions is also taken into account.
But the price to be paid
for the more accurate description of the interactions is a less complete
statistical sampling of the phase space. The simulation times are substantially
shorter than those that can be achieved in phenomenological simulations.
Longer simulations would be needed in order to fully converge the heights
of the specific heat peaks, or in order to observe a van der Waals loop in the
caloric curves, to mention two examples. But we expect that the
locations of the various transitions are reliable. All the indicators we have
used, both thermal and structural ones, are in essential agreement on
the temperature at which the transitions start.
As we discussed in a previous paper,
\cite{Agu99} longer trajectories may induce just a slight lowering in the 
transition temperatures.

The melting-like transitions of Na$_{142}$, Na$_{92}$, and Na$_{55}$ have been 
investigated by applying an orbital-free, density-functional molecular
dynamics method. The computational effort which is required is modest in 
comparison with the traditional Car-Parrinello Molecular Dynamics technique 
based on Kohn-Sham orbitals, that would be very costly for clusters of this
size. Specifically, the computational effort to update 
the electronic system scales linearly with the system size N, in contrast to 
the N$^3$ scaling of orbital-based methods. This saving allows the study of 
large clusters. However, the price to pay
is an approximate electron kinetic energy.

An icosahedral isomer of Na$_{142}$ melts in two steps as evidenced
by the thermal indicators. 
Nevertheless, there are isomerization 
transitions involving surface defects at a temperature as low as 130 K,
that preserve the icosahedral structure and do not give rise to any 
pronounced feature in the caloric curve.
The transition
at T$_s \approx$ 240 K from that isomerization regime 
to a phase in which the surface atoms aquire a substantial diffusive motion
is best described as surface melting. This is followed at T$_m 
\approx$ 270 K by homogeneous melting. 
For Na$_{92}$, there is a minor peak in C$_v$ at T$_s \approx$130K which
we associate with surface melting. The smaller value of T$_s$, for this
cluster compared with Na$_{142}$, is due to the less ordered surface.
Na$_{55}$, being a perfect two-shell icosahedron 
with no surface defects melts in a single stage at 190 K. In all cases, for
T$>$T$_m$ the atoms are able to diffuse throughout the cluster volume.
Both the calculated T$_m$ at which homogeneous melting occurs and the estimated 
latent heat of fusion q$_m$ are in excellent agreement with the experimental 
results of Haberland and coworkers for Na$_{142}$ and Na$_{92}$; our earlier 
results on the melting of Na$_8$ \cite{Agu99} are also consistent with the 
variation of the measured optical spectrum with temperature. 
A serious discrepancy between theory and experiment remains for Na$_{55}$.

We have found that structural quantities obtained from the simulations which 
are very useful in the study of melting in small clusters,\cite{Agu99} such 
as the diffusion coefficient, are not, in the case of the larger clusters
studied here, efficient indicators of homogeneous melting, which is better
located with thermal indicators. A better structural indicator is the
evolution with temperature of the average radial ion density.
This quantity flattens when homogeneous melting occurs.

$\;$

$\;$

$\;$

{\bf ACKNOWLEDGEMENTS:} We would like to thank H. Haberland, S. K\"ummel and F. Calvo
for sending us preprints of their respective works prior to publication.
This work has been supported by DGES 
(Grants PB95-0720-C02-01 and PB98-D345), NATO(Grant CRG.961128) 
and Junta de Castilla y Le\'on (VA28/99).
A. Aguado acknowledges a graduate fellowship from Junta de Castilla y Le\'on. 
M. J. Stott acknowledges the support of the NSERC of Canada and an Iberdrola
visiting professorship at the University of Valladolid.


{\bf Captions of Figures.}

{\bf Figure 1} Caloric and specific heat curves
of Na$_{142}$, taking the
internal cluster temperature as the independent variable. 
The deviation around the mean
temperature is smaller than the size of the circles.

{\bf Figure 2} Caloric and specific heat curves
of Na$_{92}$, taking the
internal cluster temperature as the independent variable. 
The deviation around the mean
temperature is smaller than the size of the circles.

{\bf Figure 3} Caloric and specific heat curves
of Na$_{55}$, taking the
internal cluster temperature as the independent variable. 
The deviation around the mean
temperature is smaller than the size of the circles.

{\bf Figure 4} Short-time averaged distances $<r_i(t)>_{sta}$ between each atom
and the center of mass in Na$_{142}$, as functions of time for 
the icosahedral isomer at T= 30 K.

{\bf Figure 5} Short-time averaged distances $<r_i(t)>_{sta}$ between each atom
and the center of mass in Na$_{142}$, 
as functions of time for the icosahedral
isomer at T= 160 K. The bold lines follow the 
evolution of a particular atom 
in the surface shell and another in the outermost core 
shell.

{\bf Figure 6} Short-time averaged distances $<r_i(t)>_{sta}$ between each atom
and the center of mass in Na$_{142}$, as functions of time at T= 361 K.
The bold lines are to guide the eye in following the diffusive behavior of 
specific atoms.

{\bf Figure 7} Time
averaged radial atomic densities of the icosahedral isomer of 
Na$_{142}$, at some representative
temperatures.

{\bf Figure 8} Diffusion coefficient as a function of temperature for the
icosahedral isomer of Na$_{142}$.

{\bf Figure 9} Calculated melting temperatures, compared with the
experimental values. The experimental values for the larger cluster 
sizes are taken from ref.  \onlinecite{Sch97}, while that for the smallest 
Na$_8$ cluster is taken from ref. \onlinecite{Sch99} (see text for details).

\onecolumn[\hsize\textwidth\columnwidth\hsize\csname
@onecolumnfalse\endcsname

\begin{figure}
\psfig{figure=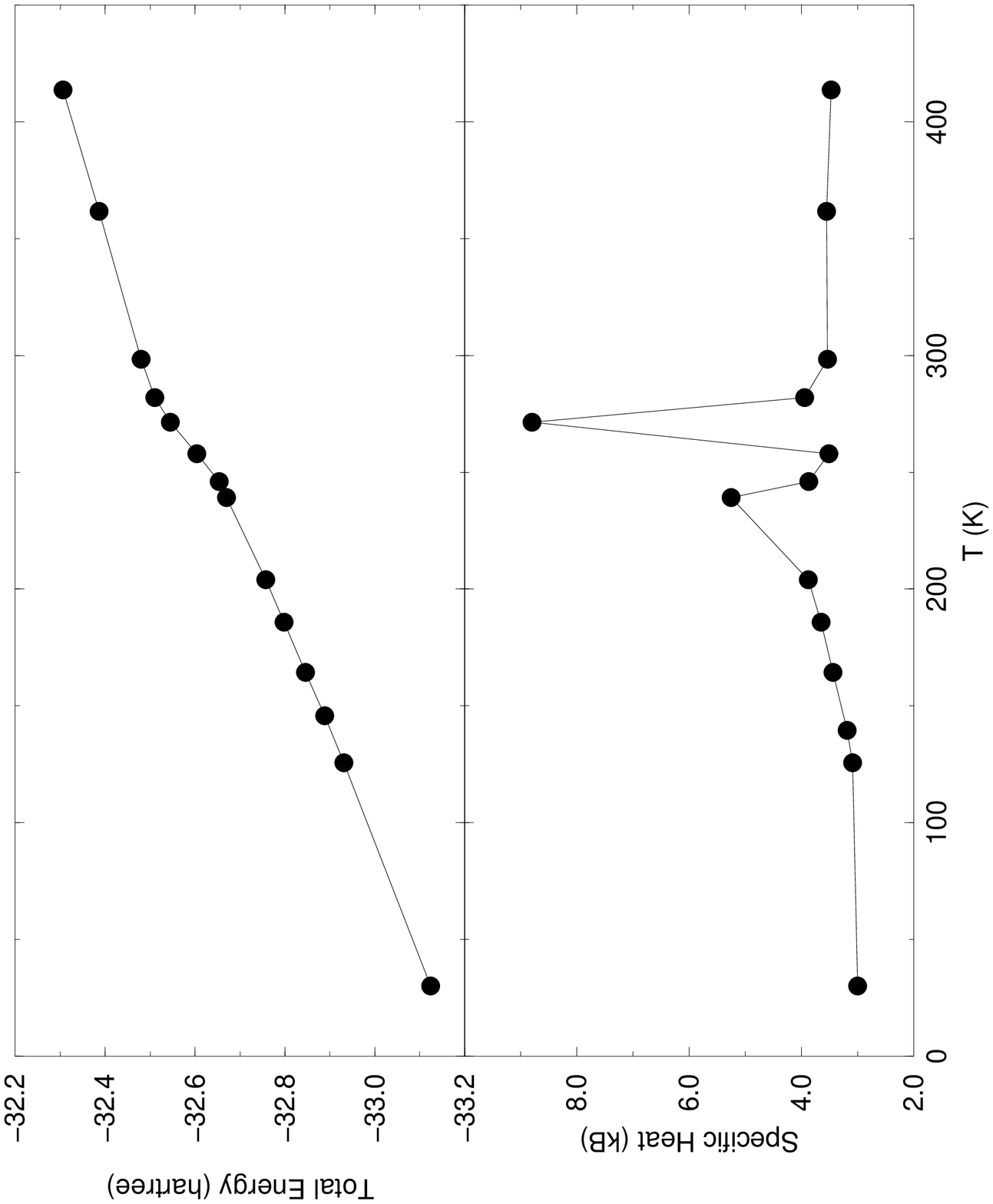}
\end{figure}

\begin{figure}
\psfig{figure=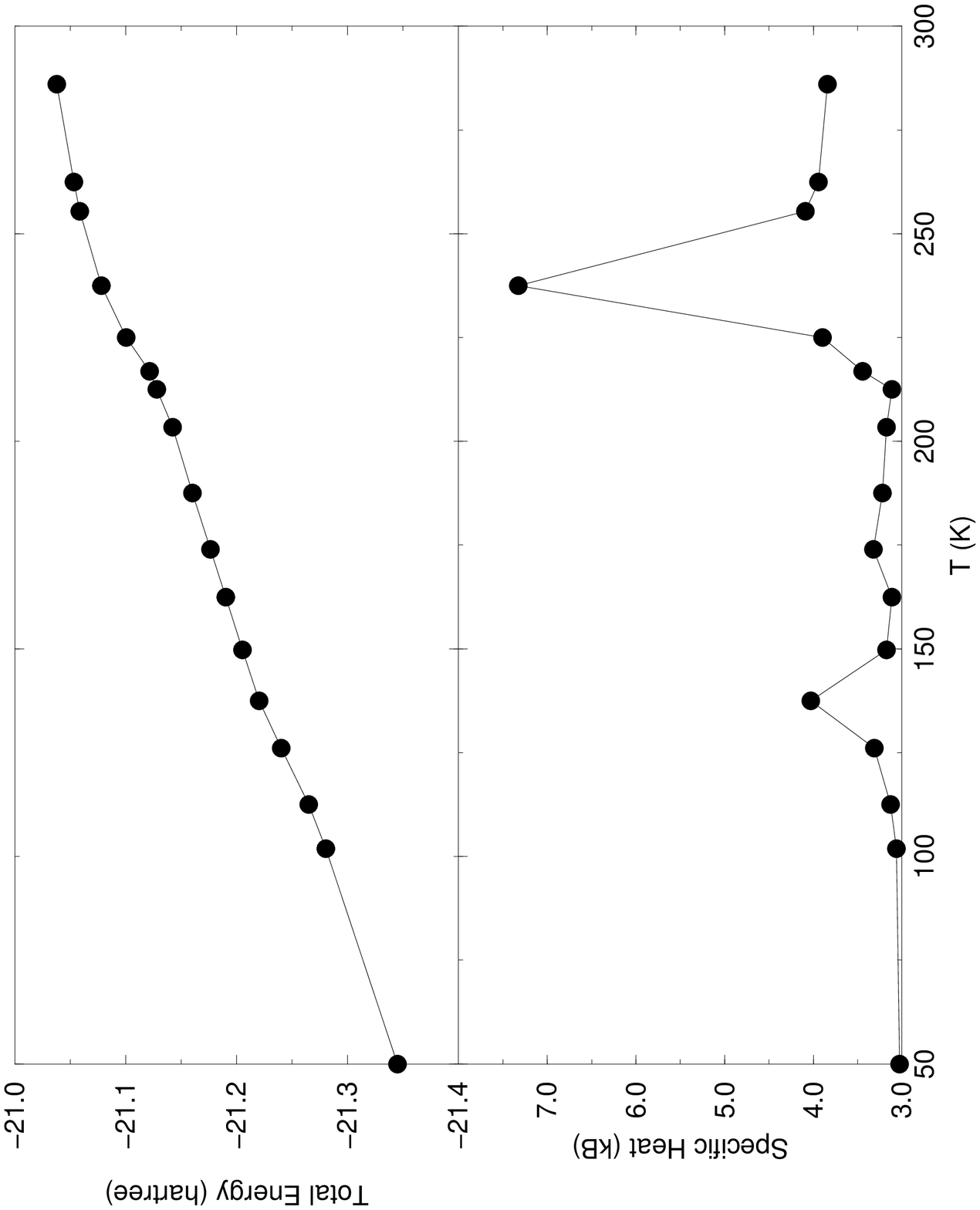}
\end{figure}

\begin{figure}
\psfig{figure=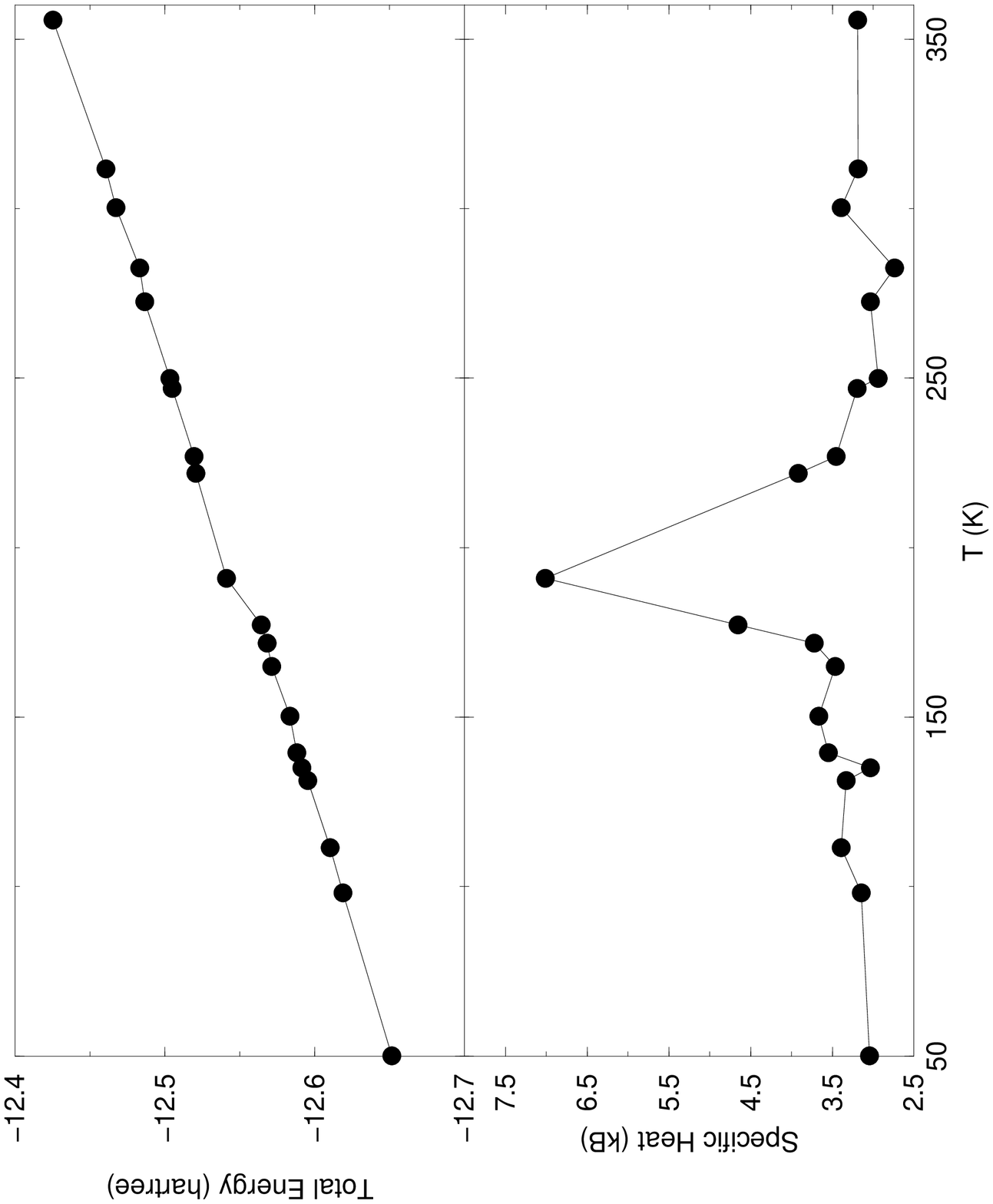}
\end{figure}

\begin{figure}
\psfig{figure=fig4a.epsi}
\end{figure}

\begin{figure}
\psfig{figure=fig5a.epsi}
\end{figure}

\begin{figure}
\psfig{figure=fig6.epsi}
\end{figure}

\begin{figure}
\psfig{figure=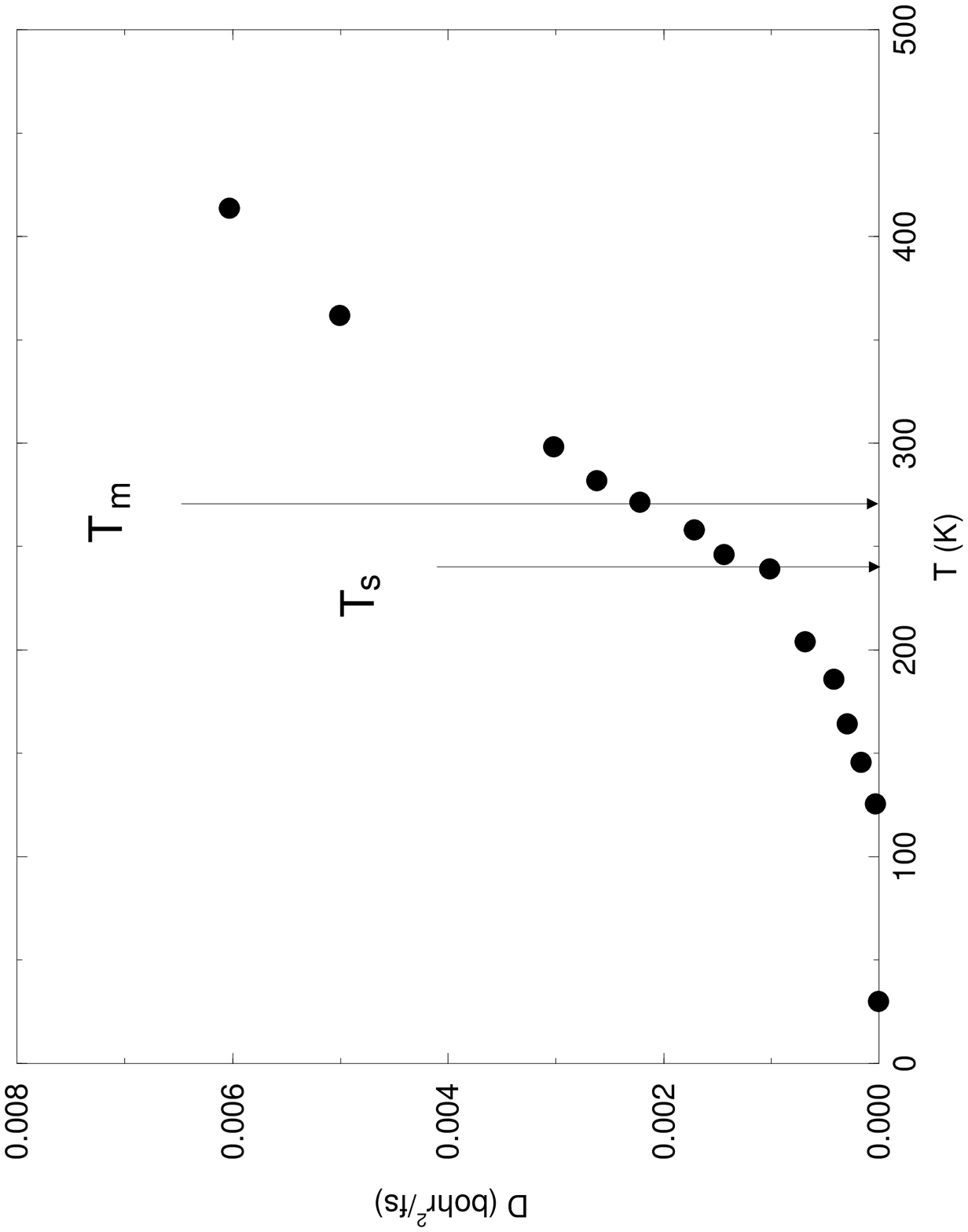}
\end{figure}

\begin{figure}
\psfig{figure=icodensity.epsi}
\end{figure}

\begin{figure}
\psfig{figure=comparison.epsi}
\end{figure}

\end{document}